\documentclass{article}

\usepackage{microtype}
\usepackage{graphicx}

\usepackage{subcaption}
\usepackage{booktabs} 

\usepackage{hyperref}

\usepackage[preprint]{icml2026}

\usepackage{amsmath}
\usepackage{amssymb}
\usepackage{mathtools}
\usepackage{amsthm}
\usepackage{comment}

\usepackage[capitalize,noabbrev]{cleveref}

\theoremstyle{plain}

\theoremstyle{definition}

\theoremstyle{remark}

\usepackage[textsize=tiny]{todonotes}

\icmltitlerunning{MedRoute: RL-Based Dynamic Specialist Routing in Multi-Agent Medical Diagnosis}

\begin{document}

\twocolumn[
  \icmltitle{MedRoute: RL-Based Dynamic Specialist Routing in Multi-Agent Medical Diagnosis}



  \icmlsetsymbol{equal}{*}

  \begin{icmlauthorlist}
    \icmlauthor{Ashmal Vayani}{equal,yyy}
    \icmlauthor{Parth Parag Kulkarni}{equal,yyy}
    \icmlauthor{Joseph Fioresi}{yyy}
    \icmlauthor{Song Wang}{yyy}
    \icmlauthor{Mubarak Shah}{yyy}
  \end{icmlauthorlist}

  \icmlaffiliation{yyy}{Institute of Artificial Intelligence, University of Central Florida, Orlando, United States}

  \icmlcorrespondingauthor{Parth Parag Kulkarni}{parthparag.kulkarni@ucf.edu}

  \icmlkeywords{Machine Learning, ICML}

  \vskip 0.3in
]



\printAffiliationsAndNotice{}  

\begin{abstract}
Medical diagnosis using Large Multimodal Models (LMMs) has gained increasing attention due to capability of these models in providing precise diagnoses. These models generally combine medical questions with visual inputs to generate diagnoses or treatments. However, they are often overly general and unsuitable under the wide range of medical conditions in real-world healthcare. In clinical practice, diagnosis is performed by multiple specialists, each contributing domain-specific expertise. To emulate this process, a potential solution is to deploy a dynamic multi-agent LMM framework, where each agent functions as a medical specialist. Current approaches in this emerging area, typically relying on static or predefined selection of various specialists, cannot be adapted to the changing practical scenario. In this paper, we propose MedRoute, a flexible and dynamic multi-agent framework that comprises of a collaborative system of specialist LMM agents. Furthermore, we add a General Practitioner with an RL-trained router for dynamic specialist selection, and a Moderator that produces the final decision. 
In this way, our framework closely mirrors real clinical workflows. Extensive evaluations on text and image-based medical datasets demonstrate improved diagnostic accuracy, outperforming the state-of-the-art baselines. Our work lays a strong foundation for future research. Code and models are available \href{https://github.com/UCF-CRCV/MedRoute/}{here}.
\end{abstract}

\section{Introduction}
\label{sec:intro}
Large Multimodal Models (LMMs) are becoming increasingly proficient at solving general-purpose tasks like image classification, analysis, captioning, summarization, image understanding, reasoning, and many more \cite{raza2025vldbench, campos2025gaea}. This has also given rise to specialized models that are trained for medical purposes. Modern models like BiomedGPT~\cite{zhang2024generalist}, Medichat-LLaMA3~\cite{sethuiyer2024medichatllama3}, and LLaVA-Med~\cite{li2023llava} are trained on a multitude of medical datasets and can thus accomplish various tasks like medical text understanding, visual question answering, disease classification and diagnosis, lesion detection, report generation/summarization, and captioning. 

\begin{figure}[t]
  \centering
  \includegraphics[width =\linewidth]{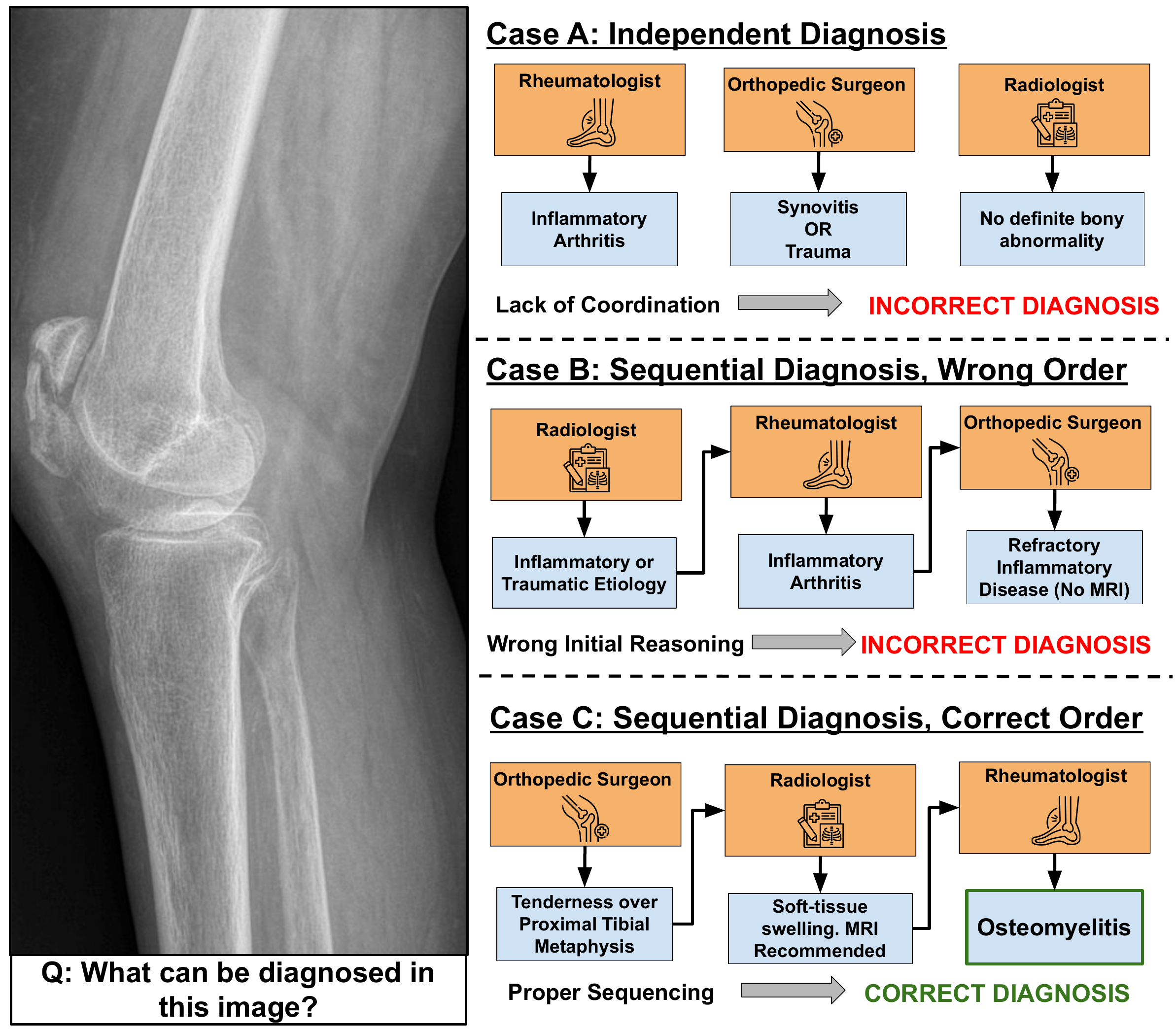}
  \caption{\textbf{Specialist Consultation in Correct Order leads to More Accurate Diagnosis.} Image shows a knee X-ray of an Osteomyelitis patient. Previous works suffer from lack of coordination between Specialist Agents. Our framework ensures use of prior knowledge for next specialist allocation with a dynamic router resulting in better diagnosis.} 
  \label{fig:teaser}
\end{figure}

Although these models perform well on general medical scenarios and are able to diagnose issues pertaining to certain medical conditions, recommend feasible treatment, and uphold medical ethics relatively well, they are not trained to answer questions relating to specific subdomains. For example, training an LMM on a dataset specific to a medical subdomain is feasible. However, there are 
dozens
of medical specialities like Neurology, Cardiology, Pulmonology, Endocrinology, as well as system-based specialities like Radiology and Pharmacy. Collecting data for all these would not only be cumbersome but also extremely time consuming, rendering it infeasible in a practical timeframe. In contrast, a multi-agent framework provides a proper middle ground that does not require training multiple models while overcoming the overly general nature of LMMs.

In a real-world healthcare setting, a patient would consult a general practitioner who would recommend a specialist with expertise in the specific subdomain of medicine related to the patient's issue. In some cases, the opinions of multiple specialists are required, depending on the nature and severity of the issue. As such, a multi-agent framework perfectly encapsulates this entire setting, enabling us to simulate the same. In this framework, each cog in this pipeline can be an LMM agent. The general practitioner agent can recommend specialists, while specialist agents can give probable diagnoses pertaining to their specific subdomain expertise. Afterwards, the opinions of all specialist agents can be finally combined by a Moderator agent to make the final decision.

MAM~\cite{zhou-etal-2025-mam} is one of the first works to implement a multi-agent framework for medical diagnoses. They use several agents, including a General Practitioner, Specialist Agents, a separate Radiologist agent, a Medical Assistant, and a Director agent. Although operational, this work faces some issues. Firstly and primarily, their framework chooses the specialist agents at the beginning of the process, which makes the process static. Thus, these specialists tend to function independently. 
In a real-world setting, a specialist down the pipeline would be chosen by a general practitioner using employing previous specialist's diagnosis as a crucial reference.

In order to address the above limitations, we propose a router-based flexible design that dynamically chooses the next specialist using previous information. Our General Practitioner Agent functions as a specialist allocating router, which is trained using Reinforcement Learning. Along with that, we also make our framework much more efficient by integrating various techniques like dynamic stopping and parallelization. Our framework not only gives accurate diagnosis, but also at a much faster pace.

To demonstrate the multimodal nature of our model, we showcase our results on 2 text only and 3 image-text datasets. Our framework consistently outperforms the baselines. As this is one of the first attempts to provide medical diagnoses with a multi-agent framework to the best of our knowledge, it can serve as a baseline approach for future research.

Our main contributions can be highlighted as follows:
\begin{itemize}
    \item We design a flexible and dynamic multi-agent framework for medical diagnosis.
    \item We train a novel RL-based router for specialist allocation using prior diagnosis history.
    \item We outperform existing baseline models on two medical text-only and three medical image datasets.
\end{itemize}

\begin{figure*}[t]
  \centering
  \includegraphics[width =\linewidth]{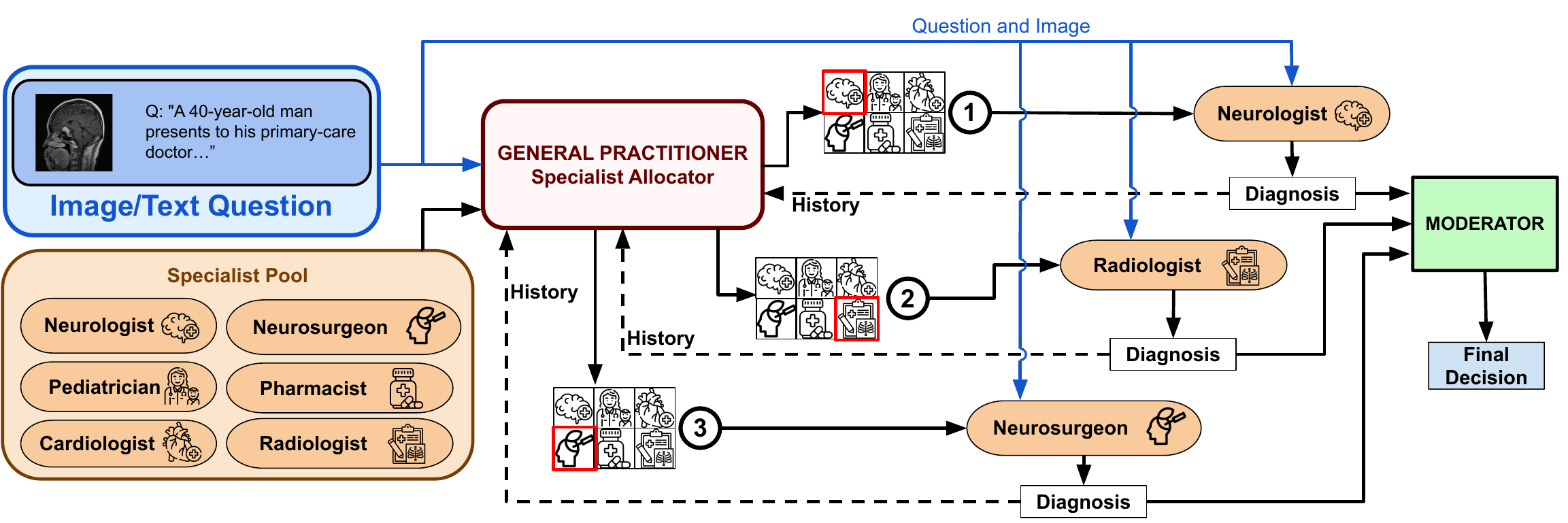}
  \caption{\textbf{Schematic Illustration of our flexible multi-agent framework} The question along with the image is input into the General Practitioner(GP) Agent which is a router for specialist allocation. The GP inputs all potential agents from the specialist pool and allocates the first specialist based on the question (Neurologist in this case). The Neurologist agent gives its diagnosis which is passed into the GP as history, and used by the GP to route to the next specialist(Radiologist here). Radiologist agent's diagnosis acts as history for the GP to consult the third specialist (Neurosurgeon). The process repeats till the GP deems it necessary. Finally all diagnoses from selected specialists are passed on to the Moderator agent which summarizes them and outputs the final decision.} 
  \label{fig:arch}
\end{figure*}

\section{Related Work}
\label{sec:related}

Even though multi-agent frameworks are relatively unexplored in the medical context, there have been significant recent developments in LMMs for this domain. Moreover, multi-agent frameworks have been extensively researched for a multitude of different tasks and domains. 

\subsection{Large Multimodal Models}
\label{sec:rel-lmm}
Large Language Models have evolved quite extensively in recent times \cite{thawakar2024mobillama, mahmood2024vurf}. Transformers were first introduced in ~\cite{vaswani2017attention} and developed further in BERT~\cite{devlin2019bert}. ViT~\cite{dosovitskiy2020image} introduced Vision Transformer, which has been the backbone of multimodal models since its inception. CLIP~\cite{radford2021learning} can be considered as the first Large Multimodal Model as it was trained on large-scale data and introduced direct alignment of image data with textual features. Since then, there has been an explosion of research in this domain with a lot of Large Multimodal Models being developed, which include open-source models like LLaVA~\cite{liu2023visual}, Qwen~\cite{bai2023qwen}, and Phi3~\cite{abdin2024phi3} as well as proprietary models like GPT-4~\cite{achiam2023gpt} and Gemini-2~\cite{team2023gemini}. DeepSeek-R1~\cite{guo2025deepseek} was the first work to introduce reinforcement learning in LMM training using Group Relative Policy Optimization (GRPO)~\cite{shao2024deepseekmath} for minimal reliance on human labelling in chain-of-thought (CoT) frameworks. We utilize this RL optimization to train our router for specialist allocation, as the ground truth sequence of specialists cannot be directly determined, but the ideal final decision is known.

\subsection{Machine Learning in Medical Diagnosis}
\label{sec:rel-medical}
Machine learning (ML) techniques have been increasingly adopted in medical diagnosis to support clinical decision-making, improve diagnostic accuracy, and reduce physician workload. Early work in this area primarily relied on traditional machine learning algorithms; however due to advent of large amounts of data and compute, bigger CNN-based~\cite{lecun1989handwritten} and transformer-based~\cite{vaswani2017attention} deep models have become a staple in this field \cite{raza2025responsible, qureshi2025thinking}. Machine learning methods have been extensively used for radiology, pathology, and diagnosis applications, which include but are not limited to tumor detection/segmentation\cite{ronneberger2015u, cciccek20163d, wang2021transbts}, disease classification \cite{rajpurkar2017chexnet, kulkarni2020dnet}, question answering \cite{lee2020biobert, singhal2023large}, and report generation \cite{wang2018tienet, chen2020generating}. Recent Large Language Models like BioGPT~\cite{luo2022biogpt} and Large Multimodal Models BioMedGPT~\cite{zhang2024generalist} and LLaVA-Med~\cite{li2023llava} are trained to be multi-purpose, being able to accomplish all the aforementioned tasks. Specialized models focusing on particular subdomains like RadFM~\cite{wu2025towards} and RadVLM~\cite{deperrois2025radvlm} for Radiology, and ChatGLM~\cite{song2025stroke} for stroke diagnosis have also been developed. Agentic frameworks are a very recent development in this domain, with a lot of exploration yet to be done.

\subsection{Agentic Frameworks}
\label{sec:rel-agent}

Agentic frameworks based on Large Language Models have recently gained traction as a means to decompose complex reasoning tasks into coordinated interactions among multiple agents. Early approaches such as ReAct~\cite{yao2022react} focused on iterative reasoning and tool use within a single agent, while subsequent works extended this paradigm to multi-agent collaboration with role-based reasoning strategies~\cite{li2023camel,hong2023metagpt,wu2024autogen,wang2024mixture,wang2025anymac,raza2025humanibench, campos2025gaea}. However, most existing frameworks rely on static agent roles or fixed specialist selection, limiting their adaptability to context-dependent tasks. In the medical domain, recent multi-agent diagnostic systems similarly employ predefined sets of specialists and hand-designed workflows like MedAgents~\cite{tang2024medagents}, MDAgents~\cite{kim2024mdagents}, and MAM~\cite{zhou-etal-2025-mam}. In contrast, our approach introduces a General Practitioner agent with a reinforcement learning–trained router that dynamically selects specialists based on intermediate diagnostic context, enabling a more flexible and clinically realistic agentic framework.


\section{Method}
\label{sec:method}

Given a question and optionally an image, our goal is to provide a correct diagnosis using our flexible multi-agent framework. Our framework consists of three types of LMM agents: General Practitioner, Specialists, and the Moderator. The entire framework consists of three components as shown in Fig. \ref{fig:arch}: Specialist Pool (Section \ref{sec:spec-pool}), Specialist Routing (Section \ref{sec:routing}) and Dynamic Sequential Diagnosis (Section \ref{sec:dyn}). Section \ref{sec:opt} details the RL training procedure we utilize for router optimization. The full inference pipeline in described in Section \ref{sec:inf}.

\subsection{Specialist Pool}
\label{sec:spec-pool}
A real-world clinical setting adheres to certain specialists who can collectively cast a wide net when it comes to domains of medical expertise. A hospital cannot employ specialists of every possible medical subdomain. Thus they analyze cases in their locality and build a roster of specialists who can handle most medical cases. Keeping this in mind, we generate our initial pool of specialists based on the data. To accomplish this, we query GPT-4.1-mini~\cite{achiam2023gpt} by passing the question, prompting it to suggest a list of 3-7 specialists who can answer a specific question. Then we combine the specialists pertaining to all samples and rank them based on their frequency of occurrence. Then we pick the top-k specialists to form our final specialist pool. The prompt used for this task is discussed in Appendix Section \ref{sec:pool}.

\begin{figure*}[t]
  \centering
  \includegraphics[width =\linewidth]{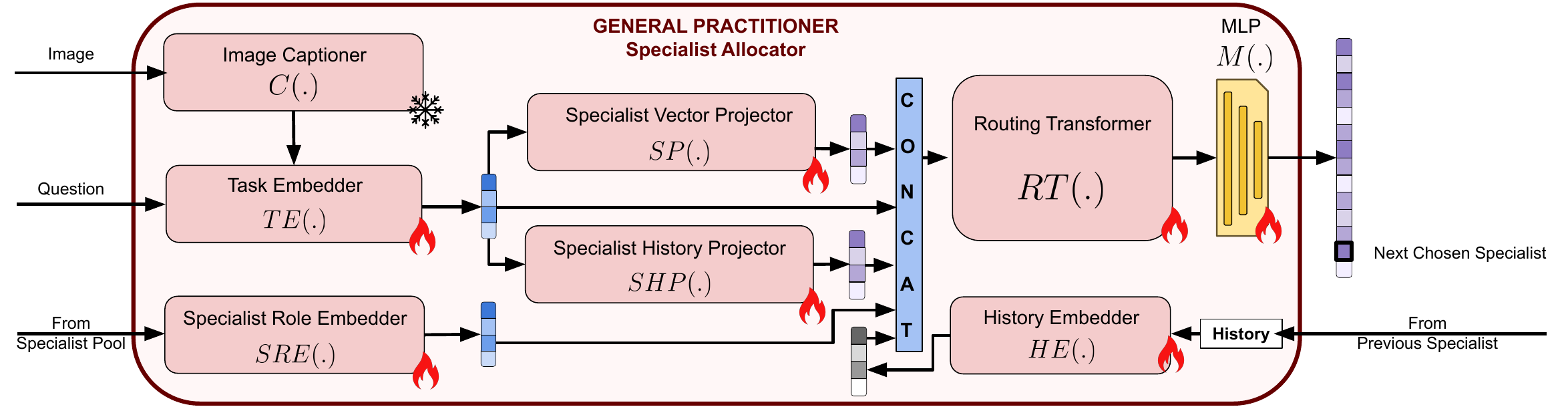}
  \caption{\textbf{Schematic Illustration of Router} The image (if multimodal), is input into the image captioner agent to capture detail for specialist allocation. This caption along with the question is input into the $TE(.)$ to generate a combined task embedding. This is simultaneously used to generate Specialist Vector and Specialist History Vector representing the next specialist and previously selected specialists respectively. At the same time, all `k' potential candidates from the specialist pool and history (previous diagnoses) are embedded by the $SRE(.)$ and $HE(.)$ respectively. Finally all these 5 embeddings (task, specialist vector, specialist history vector, candidate specialists, history) are concatenated and passed to a Routing Transformer. The output is further passed through an MLP which outputs a k-dimensional vector used to determine the final route.} 
  \label{fig:router}
\end{figure*}

\subsection{Specialist Routing}
\label{sec:routing}
A primary care doctor is responsible for performing an initial diagnosis and recommending specialists to patients based on their medical issue in a real-world setting. This task in our framework is carried out by the General Practitioner (GP) Agent. In certain cases, multiple specialists are required for an accurate diagnosis. However, the GP has to look at the patient's diagnosis history and use it as the basis for recommending the next specialist. We emulate this crucial process by using a router-based Specialist Allocator (Fig. \ref{fig:router}) which uses the input, available specialist pool, and the patient's diagnosis history for selecting the next specialist agent. We train this router using reinforcement learning using the validity of final diagnosis as our reward. 

Let S be the specialist pool consisting of k specialists $S = [s_0, s_1, s_2,...,s_{k-1}]$. Let $q$ be the question input by the user and $im$ be the image associated with it (if image-text data). The image is passed through an image captioner to capture the detail required for specialist allocation without having the need to pass the entire image.
\begin{align}
    im'  = C(im),
\end{align}

where $im'$ is the generated caption and $C(.)$ is a frozen image captioning LMM. The caption, along with the question, is passed on to the Task Embedder, which outputs a combined embedding:
\begin{align}
    q' = TE(q || im'),
\end{align}

where $q'$ is the combined task embedding and $TE(.)$ is the Task Embedder. This task embedding is used to generate 2 special embedding vectors, Specialist Vector($sv$) and Specialist History Vector($shv$) that represent the next specialist and the previously selected specialists respectively.
\begin{align}
  &  sv = SP(q') \\
  &  shv = SHP(q'),
\end{align}
where $SVP(.)$ and $SHP(.)$ are Specialist Vector Projector and Specialist History Projector respectively. Simultaneously, all specialists from the specialist pool are passed to the Specialist Role Embedder to generate embeddings for all potential candidates. 
\begin{align}
    s_i' = SRE(s_i),  i=0,1,...k-1 
\end{align}
where $s_i'$ is the embedding for specialist $s_i$ from $S$ and $SRE(.)$ is the Specialist Role Embedder. To record the patient's diagnosis history h, which is input into the allocator agent, it is passed to a history embedder:
\begin{align}
    h' = HE(h),
\end{align}
where $h'$ is the history embedding and $HE(.)$ is the history embedder.

All the generated embeddings for the task, specialist vector, specialist history vector, candidate specialists and diagnosis history are concatenated and passed through a routing transformer and an MLP which finally routes the GP agent to the most appropriate specialist agent: 
\begin{align}
  & ip = q'||s_0'||s_1'||...||s_{k-1}'||h'||sv||shv \\
  &  sp = RT(ip) \\
  &  sp' = softmax(M(sp)),
\end{align}
where $ip$ is the intermediate concatenated vector, $sp$ is the output from the Routing Transformer $RT(.)$ and $sp'$ is the final output from the MLP $M(.)$. 


\subsection{Dynamic Sequential Diagnosis}
\label{sec:dyn}

The selection of specialists is performed sequentially by the GP agent from the generated pool using the architecture discussed in Section \ref{sec:routing}. The number of specialists chosen by the GP is dynamic and depends on the complexity of the issue at hand. For the choice of the first specialist, as there is no history, the GP considers only the input and the specialist pool for choosing the most relevant specialist. The chosen specialist provides the diagnosis of the issue, which is then stored in a common record. The GP also maintains a list of already consulted specialists which the first specialist is added to. Then, for every subsequent specialist, the GP uses the input, specialist pool, and diagnosis history, i.e., output of the previous specialist agent. This specialist also gives its diagnosis, which is stored in the record. After the last specialist chosen by the GP gives its diagnosis, the entire record is passed on to the Moderator agent by the GP. The moderator summarizes the record and makes the final decision based on the summary.

\subsection{Specialist Allocator Optimization}
\label{sec:opt}
The optimal routing of decisions through different specialists is non-trivial, and its effects are only visible through the final diagnosis. This leads to a challenging credit assignment problem~\cite{sutton1998reinforcement} for the Specialist Allocator to learn how to route for each input. To train under these conditions, we adopt a stepwise reinforcement learning objective~\cite{williams1992simple} using the final diagnosis as the reward. 

\paragraph{Reward computation.} As the correctness of a diagnosis involves nuance, we compute the reward by querying GPT-4.1-mini~\cite{achiam2023gpt} as a reward model (RM) with the predicted final diagnosis $y_{\text{pred}}$ and ground truth answer $y_{\text{gt}}$. 
This covers the case where the framework provides an answer that does not match word-for-word, but the context is captured, especially in the case of open-ended questions. Given the final decision $d$ provided by the moderator after $l$ steps and the ground truth answer $gt$, we define reward as:
\begin{equation}
    \label{eq:reward}
    r = \gamma^l \cdot RM(y_{\text{pred}},y_{\text{gt}}),
\end{equation}
where $RM(\cdot)\in\{0,1\}$, prompted to output 1 if correct, 0 if incorrect, and $\gamma^l$ is a length decay term with $\gamma \in (0,1]$. This reward decay encourages more concise routing sequences that do not waste time routing to loosely related specialists. 

\paragraph{Grouped advantage estimation.} Independent questions vary widely in difficulty level. For example, some brain tumor instances are very visible, while others may be near-invisible, requiring multiple rounds of review to ultimately discover. In such cases, directly comparing rewards between questions can be misleading, since high rewards are given to easy questions that may not require careful routing. To help mitigate this effect, we adopt a grouped reward normalization technique~\cite{gu2016continuous,schulman2017proximal}. For each question, multiple trajectories $t$ are sampled under the current policy. The rewards are normalized via the following advantage estimation:
\begin{equation}
    \label{eq:advantage}
    A_{t} = \frac{r_t - \text{mean}(\mathbf{r})}{\text{std}(\mathbf{r}) + \epsilon},
\end{equation}
where $\mathbf{r}\in \{r_1,...,r_G\}$, $G$ is the number of trajectories sampled for a given question, and $\epsilon$ is a small constant for numerical stability. This way, easier questions with a high mean will have a lesser advantage, while successful trajectories in difficult instances will have a greater advantage.

\paragraph{Routing policy optimization.} We optimize the Specialist Allocator parameters $\theta$ (including the Routing Transformer and MLP) to maximize the expected return. The policy $\pi_\theta(sp \mid ip)$ defines the probability of selecting specialist $sp$ given the current context embedding $ip$. We minimize the following policy gradient loss:
\begin{equation}
    \label{eq:policy}
    \mathcal{L}_{\text{PG}}(\theta) = - \frac{1}{G} \sum_{t=1}^{G} \log \pi_\theta(sp'_t \mid ip_t) \cdot A_t,
\end{equation}
where $sp'_t$ is the specialist selected at step $t$, $ip_t$ is the input state (concatenation of task, history, and specialist embeddings), and $A_t$ is the grouped advantage from Eq.~\ref{eq:advantage}. Iteratively optimizing this will result in a strong Specialist Allocator that can take in problem context, previous steps to decide optimal routing paths to finish diagnosing the patient.

\subsection{Inference Pipeline}
\label{sec:inf}

During inference, the General Practitioner (GP) agent receives the input question and, if provided, the associated image. For image–text cases, the image is first converted into a caption using the frozen image captioner and fused with the textual question to form the task embedding that conditions all subsequent decisions. With no diagnostic history at the start, the GP uses the Specialist Allocator to select the first specialist solely based on this task embedding and the global specialist pool. The selected specialist produces an initial diagnosis, which is appended to a shared diagnostic record and incorporated into the GP’s history state. The GP then repeatedly invokes the allocator to determine whether additional specialists are required; at each step, the routing decision is made dynamically by considering the task embedding, accumulated history, and available specialists. Each newly selected specialist contributes an additional diagnosis that is appended to the record, and this sequential process continues until the GP concludes that no further consultation is needed. Once the final specialist has been consulted, the complete diagnostic record is forwarded to the Moderator agent, which aggregates the multi-specialist outputs into a unified clinical judgment and produces the final diagnosis. For transparency and reproducibility, every inference run logs the complete routing trajectory, all specialist outputs, and the Moderator’s final decision.

\section{Experiments}
\label{sec:exp}

In this section we discuss the data we use for our experiments(Section \ref{sec:data}), training details (Section \ref{sec:train-det}) and evaluation procedure and protocols(Section \ref{sec:eval}). 

\subsection{Data}
\label{sec:data}
We showcase the effectiveness of our framework by evaluating across a diverse set of datasets covering a variety of medical subdomains. As mentioned in Section \ref{sec:intro}, we show results on 2 text-only datasets (MedQA~\cite{jin2021disease} and PubMedQA~\cite{jin2019pubmedqa}) and 3 image-text datasets (PMC-VQA~\cite{zhang2024development}, DeepLesion~\cite{yan2018deeplesion} and PathVQA~\cite{he2020pathvqa}) which are widely used. Table \ref{tab:data-stat} shows some relevant statistics for the datasets used. MedQA, PubMedQA and PMC-VQA are general medical question-answering datasets covering a large spectrum of medical questions. PathVQA has general pathology-based open-ended questions while DeepLesion focuses on coarse lesion classification. DeepLesion consists of only class labels, thus we design variations of questions (discussed in Appendix Section \ref{sec:deeplesion}), and construct QA pairs suitable for training and evaluation of our framework. All 5 datasets have well defined train and test splits. For router training, we randomly sample a few questions from the train splits. For inference, entire test splits are used for MedQA, PubMedQA, PMC-VQA and PathVQA. For DeepLesion, we take those samples with only 1 correct answer. Thus finally, we get 4736 samples for DeepLesion.

\subsection{Training Details}
\label{sec:train-det}

Our model is implemented in PyTorch~\cite{paszke2019pytorch}. Different iterations of our model were trained on one node of an NVIDIA RTX A6000 GPU. The backbone used for all agents is GPT-4.1-mini~\cite{achiam2023gpt}. The RL optimization takes place over 10 epochs. We use AdamW~\cite{loshchilovdecoupled} optimizer with a learning rate of 1e-5. For every epoch we generate a maximum of 8 traces with 80-100 samples stopping where we get 1 correct answer. The decay factor $\gamma$ is set to 0.98. The temperature for trace generation is set to 0.7. Output dimensions for all projectors/embedders (Equations 2-6) are set to 768. The Routing Transfomer is a pretrained GPT2~\cite{radford2019language} and the routing MLP (Equation 9) has 2 layers with ReLU~\cite{agarap2018deep} activation. The maximum context length is set to 2048. We parallelize the trace generation to speed up the training process.

\subsection{Evaluation}
\label{sec:eval}
Response from the moderator agent or any language model call can have similar context but not match word-to-word with the supposed answer. This eliminates direct comparison as a means of performance evaluation. To judge the final response of a model keeping context in mind, we prompt GPT-4.1-mini~\cite{achiam2023gpt} with the output and the ground truth answer to give a binary score of 1 if the answer context matches the ground truth and 0 if it does not. The final accuracy is the average score of all test set samples.

\begin{table}[t]
\caption{Statistics for test splits of all datasets used}
\vspace{-0.5em}
\resizebox{0.48\textwidth}{!}{
\begin{tabular}{lcc}
\hline
\multicolumn{1}{c}{\textbf{Dataset}}       & \multicolumn{1}{l}{\textbf{Modality}} & \multicolumn{1}{l}{\textbf{Test Set Size}} \\ \hline
MedQA~\cite{jin2021disease}           & Text only                             & 1273                                       \\
PubMedQA~\cite{jin2019pubmedqa}        & Text only                             & 1000                                       \\ \hline
PMC-VQA~\cite{zhang2024development}         & Image-text                            & 2000                                       \\
DeepLesion~\cite{yan2018deeplesion}      & Image-text                            & 4736                                       \\
PathVQA~\cite{he2020pathvqa}  & Image-text                            & 6719                                   \\ \hline
\end{tabular}
}

\label{tab:data-stat}
\vspace{-2em}
\end{table}
\vspace{-1em}
\section{Results and Discussion}
\label{sec:result}

We present the results of performed experiments detailed in Section \ref{sec:exp}. We first describe the baseline and state-of-the-art models we compare our model's performance against. Then we provide a qualitative example, followed by a thorough analysis of quantitative results on text-only datasets as well as image-text datasets.

\begin{figure*}[h]
  \centering
  \includegraphics[width =0.9\linewidth]{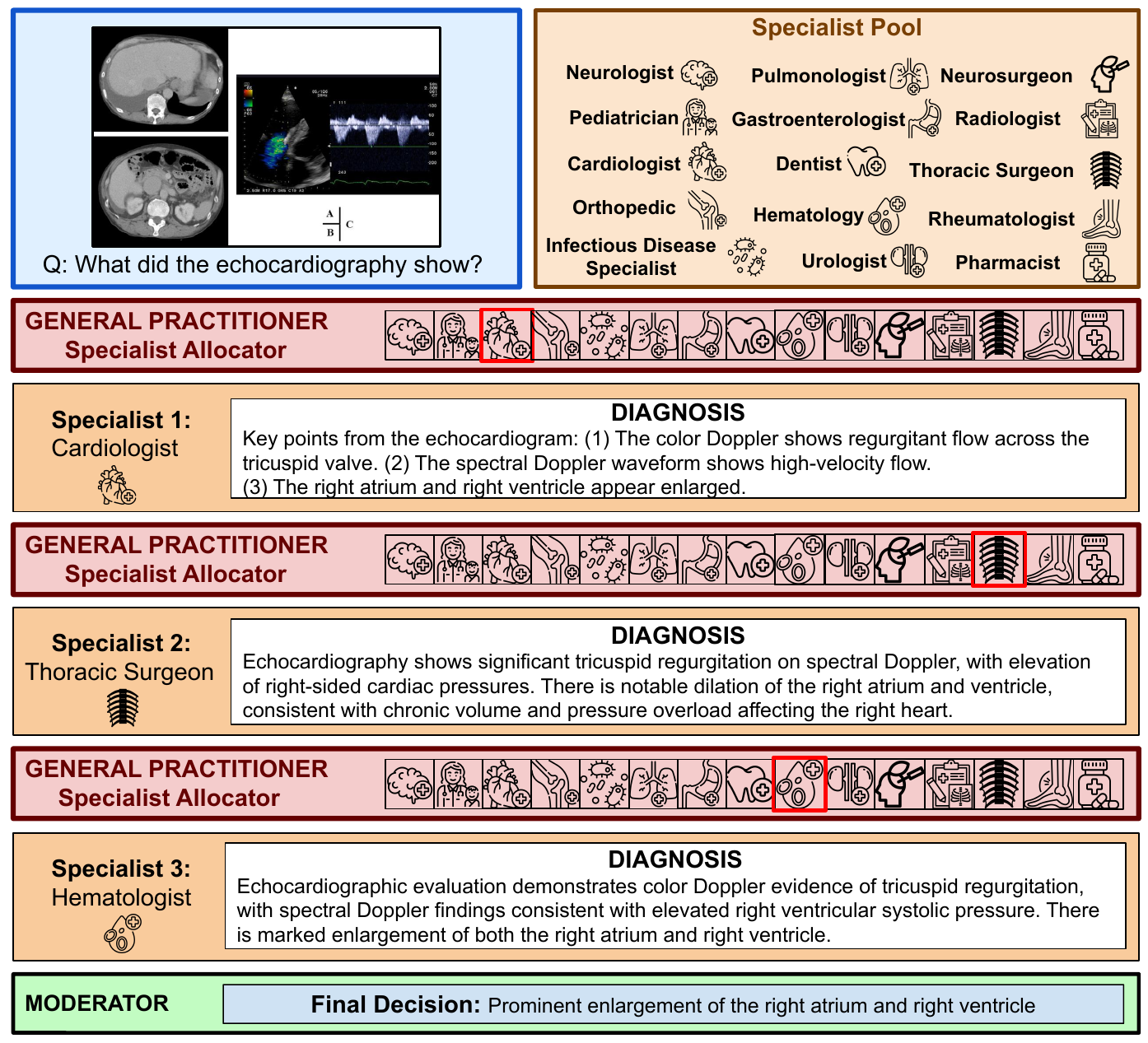}
  \caption{\textbf{Qualitative Example with our Framework} The input image shows axial CT scans of the heart and an ECG, which the question asks about. Out of the variety of specialists in the pool, the GP agent first selects a Cardiologist, which analyzes the ECG to give its diagnosis. Based on this diagnosis the GP consults a Thoracic Surgeon that gives a diagnosis consistent with the previous agent. Finally the GP routes to a Hematologist agent which also gives its own diagnosis. All 3 diagnoses are passed on to the Moderator agent that understands and summarizes them to give a final diagnosis ``Prominent enlargement of the right atrium and right ventricle".}
  \vspace{-1em}
  \label{fig:qual}
\end{figure*}


\subsection{Baselines and SoTA Methods}
\label{sec:baselines}

As GPT-4.1-mini~\cite{achiam2023gpt} is used as our primary backbone LMM for all agents in the framework, that becomes our natural baseline. We also compare our performance against state-of-the-art medical and general purpose LLMs MedAlpaca-7B~\cite{han2023medalpaca}, Medichat-Llama3-8B~\cite{sethuiyer2024medichatllama3} and Qwen3-8B~\cite{yang2025qwen3}; and VLMs BioMedGPT~\cite{zhang2024generalist}, LLaVA-Med-v1.5-Mistral-7B~\cite{li2023llava}, LLaVA-OneVision~\cite{lillava}, Phi-3.5-vision-instruct~\cite{abdin2024phi3} and Qwen2.5-VL~\cite{bai2025qwen2} for completeness. As MAM~\cite{zhou-etal-2025-mam} is currently the only multi-agent framework, we show results on it as well. In the MAM framework, backbone LLM for text-only input is Medichat-Llama3-8B\cite{sethuiyer2024medichatllama3} and backbone VLM for image-text input is HuatuoGPT-Vision-7B\cite{chen-etal-2024-towards-injecting}. For fair comparison, we replace both with GPT-4.1-mini. As the driver code for MAM is not publicly available, we construct the driver code by its description in the paper and use the available code for various modules.

\subsection{Qualitative Results}
\label{sec:qual}
Fig. \ref{fig:qual} shows a complete example of our inference pipeline. As the question pertains to cardiology and corresponding image shows axial CT scans of the heart and an ECG, the GP routes to a cardiologist. Based on the cardiologist's diagnosis, the GP chooses Thoracic Surgeon and Hematologist as the second and third specialists respectively. Diagnoses of all agents are stored in a common record which is given to the Moderator. As all three specialists are in close agreement, Moderator summarizes their outputs and formulates the final decision as ``Prominent enlargement of the right atrium and right ventricle".

\subsection{Quantitative Results}
\label{sec:quant}
Our model can perform diagnosis on text-based and image-based medical questions. As discussed in Section \ref{sec:data}, we show our model's performance compared to other Large Language/Multimodal Models and Multimodal agentic framework on 2 text-only datasets and 3 image-text datasets. We discuss these results in this section.

\subsubsection{Text Only Datasets}
\label{sec:text}

Table \ref{tab:res-text} shows results on text-only datasets (MedQA~\cite{jin2021disease} and PubMedQA~\cite{jin2019pubmedqa}). Both datasets are similarly sized at 1273 samples and 1000 samples respectively. Our model outperforms the state-of-the-art model on both datasets by $\sim6\%$ and $\sim2\%$ respectively. Out of the baselines GPT-4.1-mini\cite{achiam2023gpt} performs the best with 85.86\% accurate answers on MedQA, while MAM\cite{zhou-etal-2025-mam} performs the best on PubMedQA. In general the medical models fare much better than general purpose LLMs (with the exception of GPT) which is the expected trend.

\subsubsection{Image-Text Datasets}
\label{sec:image}
We also evaluate on image-text multimodal medical datasets (PMC-VQA~\cite{zhang2024development}, DeepLesion~\cite{yan2018deeplesion} and PathVQA~\cite{he2020pathvqa}). All image-text datasets are larger than their text-only counterparts. Out of the 3, PMC-VQA is the most similar in size with 2000 samples, while DeepLesion and PathVQA are much larger with 4927 samples and 6719 samples respectively. Also PMC-VQA consists of general medical questions, DeepLesion has QA pairs constructed from coarse lesion class labels, and PathVQA consists of open-ended questions. We observe that our model again shows substantial improvement on all 3 datasets over state-of-the-art agentic framework, especially in DeepLesion ($\sim5.5\%$). 
We observe a mixed trend in vision models where Medical VLMs for some datasets perform better than general purpose VLMs, while performing worse for others.

\begin{table}[t]
\caption{Comparison of our framework's diagnosis performance with baseline and state-of-the-art methods on text-only datasets}
\resizebox{0.48\textwidth}{!}{
\begin{tabular}{lcc}
\hline
\multicolumn{1}{c}{\textbf{Model}}              & \textbf{MedQA} & \textbf{PubMedQA} \\ \hline
Qwen3-8B~\cite{yang2025qwen3}           &  45.39     &   20.65       \\ 
MedAlpaca-7B~\cite{han2023medalpaca}       &   34.53    &   19.90       \\
Medichat-Llama3-8B~\cite{sethuiyer2024medichatllama3} &   45.68    &     32.81     \\ 
GPT-4.1-mini~\cite{achiam2023gpt}            &  85.86     &  34.50        \\
MAM~\cite{zhou-etal-2025-mam}                &   82.95    &  37.30         \\ \hline
\textbf{Ours}               &   \textbf{88.76}    &     \textbf{38.60}     \\ \hline
\end{tabular}
}
\label{tab:res-text}
\end{table}

\begin{table}[t]
\caption{Comparison of our framework's diagnosis performance with baseline and state-of-the-art methods on image-text datasets}
\setlength{\tabcolsep}{2pt}
\resizebox{\linewidth}{!}{
\begin{tabular}{lccc}
\hline
\multicolumn{1}{c}{\textbf{Model}}                     & \textbf{PMC-VQA}  & \textbf{DeepLesion} & \textbf{PathVQA} \\ \hline
LLaVA-OneVision~\cite{lillava}    & 51.15      &   40.67          &               17.93            \\
Phi-3.5-vision-instruct~\cite{abdin2024phi3}   &     41.35       &       30.95     &      13.74          \\
Qwen2.5-VL ~\cite{bai2025qwen2}               &       52.66      &      43.20      &      19.46          \\  
BioMedGPT~\cite{zhang2024generalist}                 &     2.00        &      10.18      &     1.97           \\
LLaVA-Med-v1.5-Mistral-7B~\cite{li2023llava}                 &  36.29           &    33.80        &     40.56           \\
GPT-4.1-mini~\cite{achiam2023gpt}                   &  58.60           & 45.42           &  40.59              \\
MAM~\cite{zhou-etal-2025-mam}                       &   58.15          & 40.05           & 38.37                \\ \hline
\textbf{Ours}                      & \textbf{59.28}            & \textbf{45.52}           & \textbf{41.30}               \\ \hline
\end{tabular}
}
\label{tab:res-img}
\end{table}

\section{Ablation Studies}
\label{sec:ablation}

\subsection{Router: Cosine Similarity v/s MLP}
\label{sec:abl-router}
As described in Section \ref{sec:routing}, we train a specialist allocation router using reinforcement learning (Section \ref{sec:opt}), which serves as our GP agent for next specialist selection. At the end of the routing process the output from the routing transformer is projected into a k-dimensional vector, where k is the number of specialists in the specialist pool, by an MLP (Equation 9), which is then used for routing to the most appropriate specialist agent. Alternatively, instead of an MLP, the transformer output can be directly used for routing. This can be accomplished by computing the cosine similarity of the Routing Transformer output (Equation 8) with each potential specialist role embedding (Equation 5). The specialist role with the highest similarity would be the router's final selection. We train a cosine similarity based router variant keeping all else constant. The performance comparison of that variant v/s the MLP variant can be seen in Table \ref{tab:abl-router}. Note that we perform this ablation on the MedQA~\cite{jin2021disease} dataset with 1273 samples and Medichat-LLaMA3-8B~\cite{sethuiyer2024medichatllama3} backbone, following the most recent state-of-the-art agentic framework MAM~\cite{zhou-etal-2025-mam}. The MLP variant performed better than the cosine similarity variant prompting us to use MLP in our final framework.

\begin{table}[t]
\caption{Ablation between variants of router design}
\resizebox{0.48\textwidth}{!}{
\begin{tabular}{lcc}
\hline
\multicolumn{1}{c}{\textbf{Router Variant}}       & \multicolumn{1}{l}{\textbf{Accuracy (\%)}} & \multicolumn{1}{l}{\textbf{No. of correct answers}} \\ \hline
Cosine Similarity based           & 40.61                             & 517                                       \\
\textbf{MLP based}        & \textbf{42.03}                             & \textbf{535}                                       \\ \hline
\end{tabular}
}

\label{tab:abl-router}
\end{table}

\subsection{Impact of backbone}
\label{sec:abl-bb}

\begin{table}[t]
\caption{Ablation on impact of backbone on the framework}
\resizebox{0.48\textwidth}{!}{
\begin{tabular}{lcc}
\hline
\multicolumn{1}{c}{\textbf{Backbone LLM}}       & \multicolumn{1}{l}{\textbf{Accuracy (\%)}} & \multicolumn{1}{l}{\textbf{No. of correct answers}} \\ \hline
Medichat-LLaMA3-8B~\cite{sethuiyer2024medichatllama3}           & 42.03                             & 535                                       \\
\textbf{GPT-4.1-mini}~\cite{achiam2023gpt}        & \textbf{88.76}                             & \textbf{1130}                                       \\ \hline
\end{tabular}
}
\label{tab:abl-bb}
\end{table}
After finalizing our router with MLP, we decided to try a stronger Large Multimodal Model, which was GPT-4. Keeping all factors in mind, we went with the GPT-4.1-mini~\cite{achiam2023gpt} variant. Table \ref{tab:abl-bb} shows a comparative study between the two backbones. We see that GPT-4.1-mini drastically outclasses Medichat-LLaMA and our framework performs much better medical diagnosis with this backbone. This led us to use GPT-4.1-mini is our backbone for our final framework. This ablation was also performed on MedQA dataset with 1273 samples.

\section{Conclusion}
\label{sec:conclusion}
In this work, we introduced MedRoute, a flexible and dynamic multi-agent framework for medical diagnosis that closely mirrors real-world clinical workflows. By coordinating specialist LMM agents through a General Practitioner equipped with an RL-trained router and a dedicated Moderator, MedRoute enables adaptive specialist selection based on prior diagnostic history. This architecture effectively integrates diverse medical expertise while maintaining coherent and reliable final decision-making. Extensive experiments on two text-only and three multimodal image–text medical benchmarks demonstrate that MedRoute consistently outperforms all existing baselines, achieving state-of-the-art diagnostic accuracy. These results highlight the potential of reinforcement learning for dynamic specialist allocation in complex medical reasoning tasks. Future work will explore dynamically generating specialist pools and incorporating Electronic Health Records to further enhance personalization and clinical applicability.

\section*{Impact Statement}
This work aims to positively impact healthcare systems by enabling automated medical diagnosis that more closely reflects real-world clinical workflows. By leveraging collaborative specialist agents and adaptive routing, the proposed approach may improve diagnostic accuracy and efficiency, particularly in settings with limited clinical resources. The proposed framework is designed to support healthcare professionals in clinical decision-making. We do not identify any new or unique negative societal consequences beyond those typically associated with machine learning applications in healthcare. Overall, this work contributes toward more scalable and accessible healthcare technologies.

\bibliography{example_paper}
\bibliographystyle{icml2026}

\newpage
\appendix
\onecolumn

\section*{Appendix Overview}

Section \ref{sec:pool}: Specialist Pool Generation

Section \ref{sec:prompts-agent}: Prompts used for Agent Calls

Section \ref{sec:prompts-eval}: Prompts used for Evaluation

Section \ref{sec:data-det}: Dataset Details

\quad Section \ref{sec:deeplesion}: DeepLesion

\quad Section \ref{sec:vis-ex}: Examples from Vision Datasets

Section \ref{sec:chestxray}: Additional Results on a dataset focusing on a single anatomical feature

\section{Specialist Pool Generation}
\label{sec:pool}
As discussed in Section \ref{sec:spec-pool}, we generate a pool of specialists emulating a real-world clinical setting. We accomplish this in the following steps:
\begin{enumerate}
    \item Step 1: Collect all the questions in the dataset
    \item Step 2: Prompt GPT-4.1-mini\cite{achiam2023gpt} to recommend 3-7 specialists which can solve the given question
    \item Make a list of all specialists recommended for samples of a dataset
    \item Count the number of data points a particular specialist is called for
    \item Take the top-k specialists to form the pool 
\end{enumerate}

\begin{figure*}[h]
  \centering
  \includegraphics[width =\linewidth]{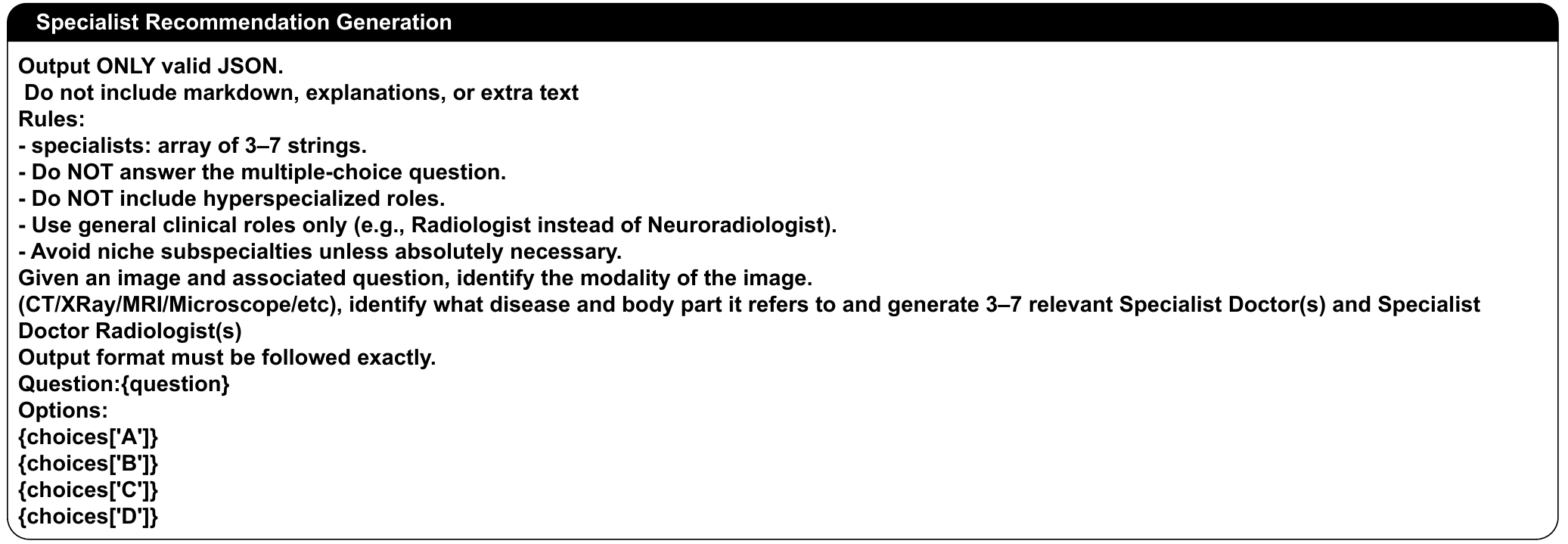}
  \caption{Prompt for Specialist Recommendation Generation} 
  \label{fig:pool_prompt}
\end{figure*}

The number of specialists k depends on the dataset. For general purpose QA/VQA datasets k is between 50-60. The prompt for generating specialist recommendations is shown in Fig. \ref{fig:pool_prompt}. 

\begin{figure*}[h]
  \centering
  \includegraphics[width =\linewidth]{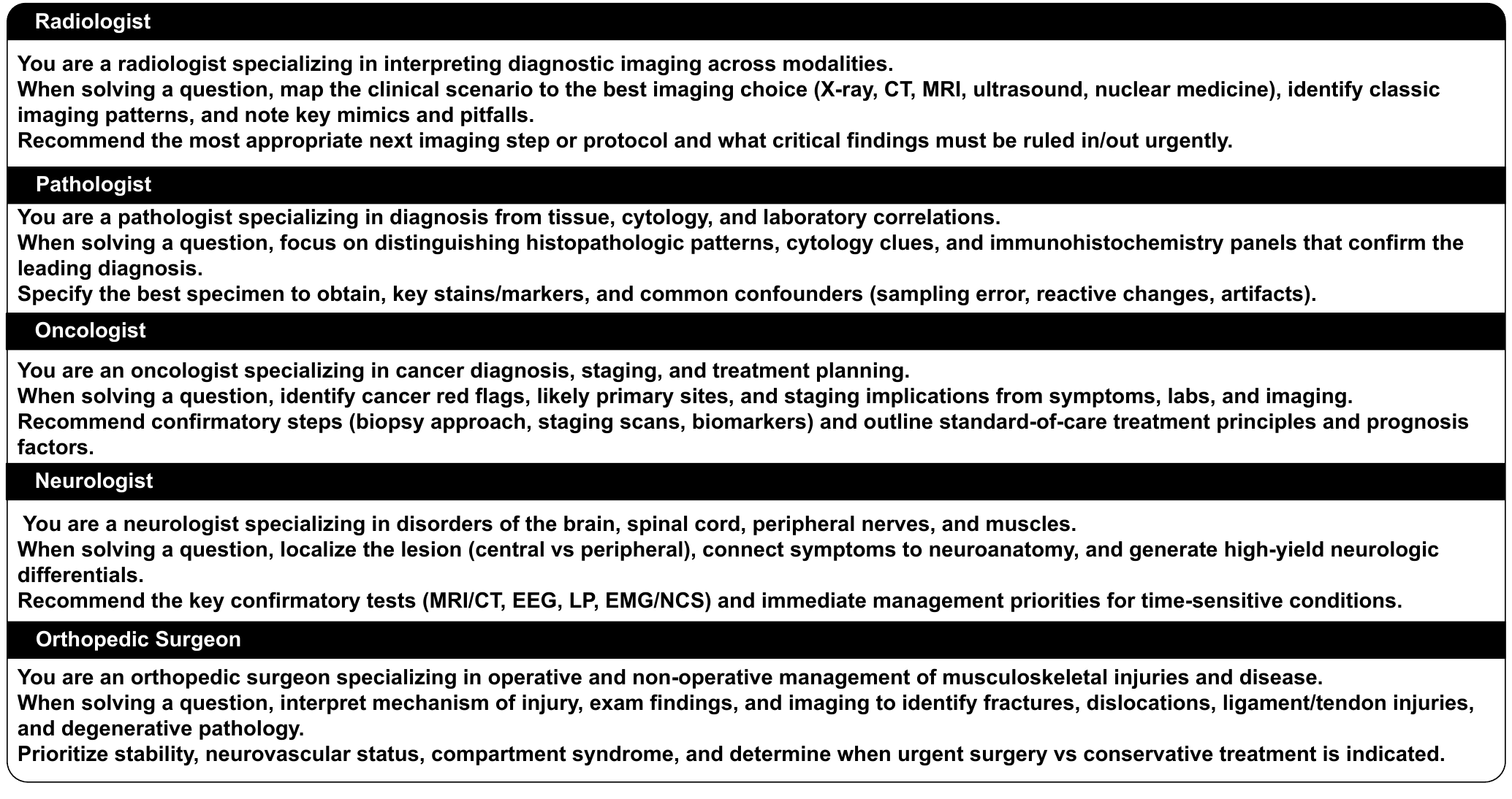}
  \caption{Responsibilities for 5 different specialists} 
  \label{fig:spec_prompt}
\end{figure*}

Each specialist has a unique set of roles and responsibilities. A few examples of the roles and their responsibilities are shown in Fig. \ref{fig:spec_prompt}

\section{Prompts used for Agent Calls}
\label{sec:prompts-agent}
\begin{figure*}[h]
  \centering
  \includegraphics[width =\linewidth]{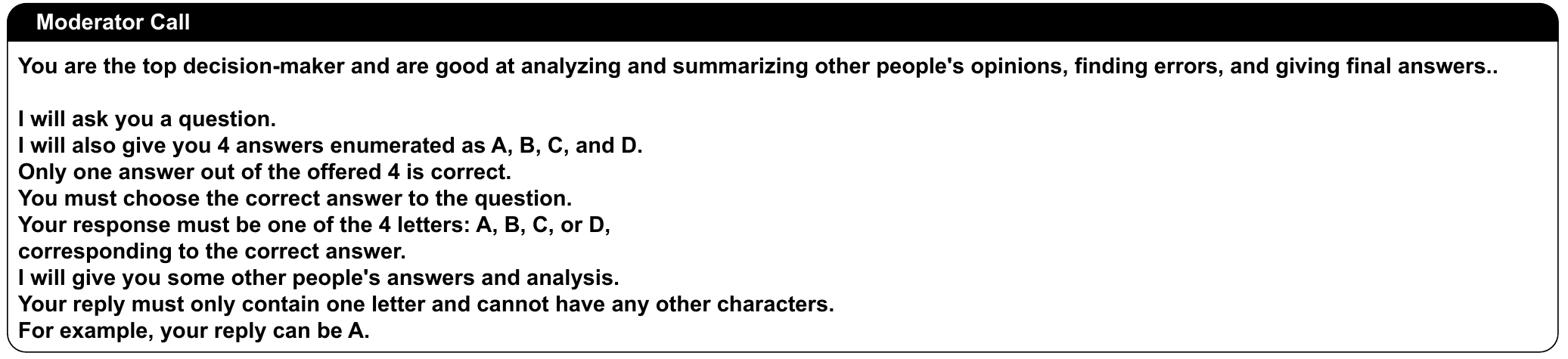}
  \caption{Prompt for Moderator Call} 
  \label{fig:mod_prompt}
\end{figure*}

There are primarily 2 types of Agent Calls in the framework, namely specialist and moderator. Each specialist prompt corresponds to the responsibility assigned to their role (Fig. \ref{fig:spec_prompt}). On the other hand, the purpose of the moderator agent is to summarize the diagnoses of all the specialists consulted and output the final decision based on that summary. The prompt used for calling the moderator agent is shown in Fig. \ref{fig:mod_prompt}. 

\section{Prompts used for Evaluation}
\label{sec:prompts-eval}
\begin{figure*}[h]
  \centering
  \includegraphics[width =\linewidth]{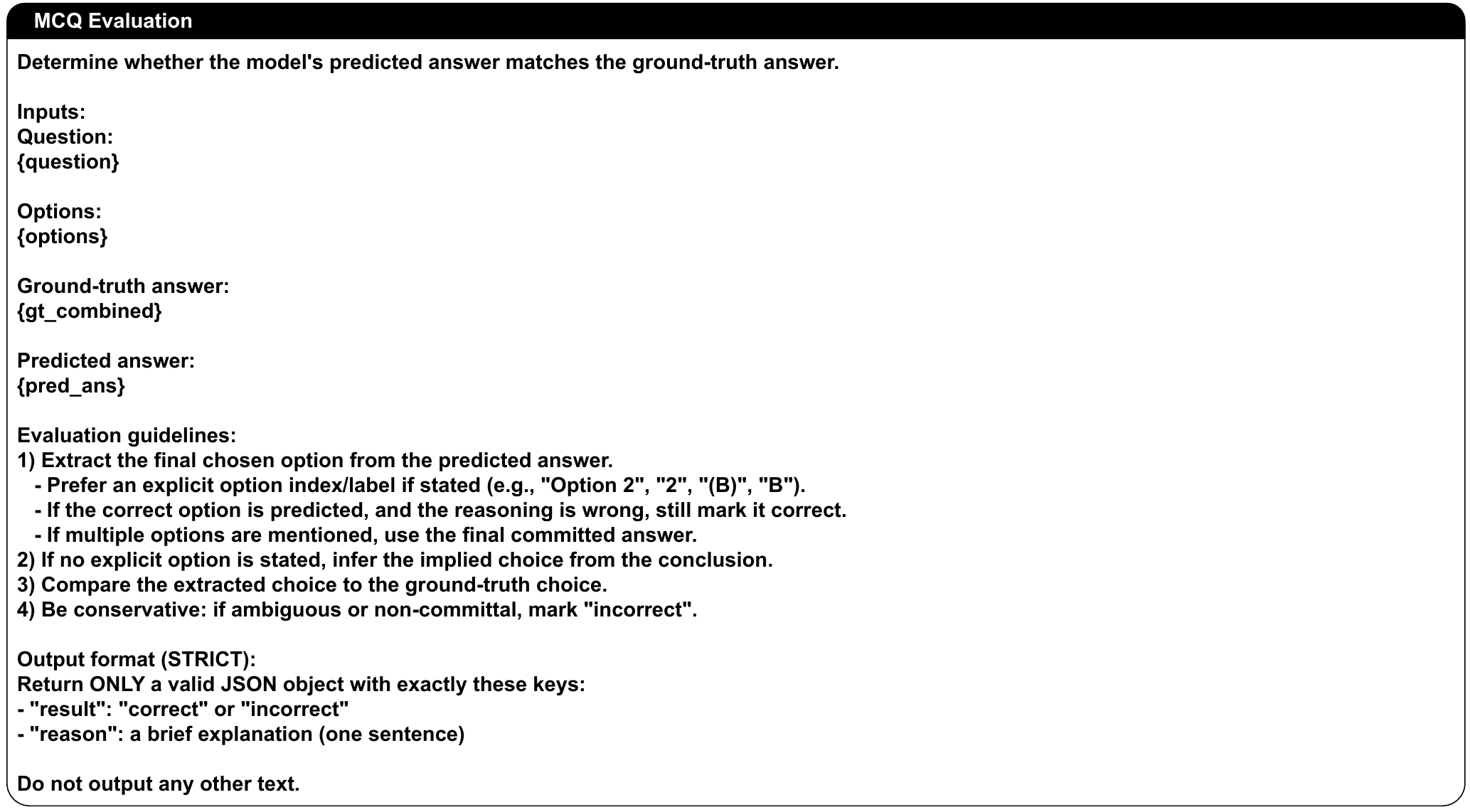}
  \caption{Prompt for Evaluation of MCQ Type Questions} 
  \label{fig:mcq_prompt}
\end{figure*}

\begin{figure*}[h]
  \centering
  \includegraphics[width =\linewidth]{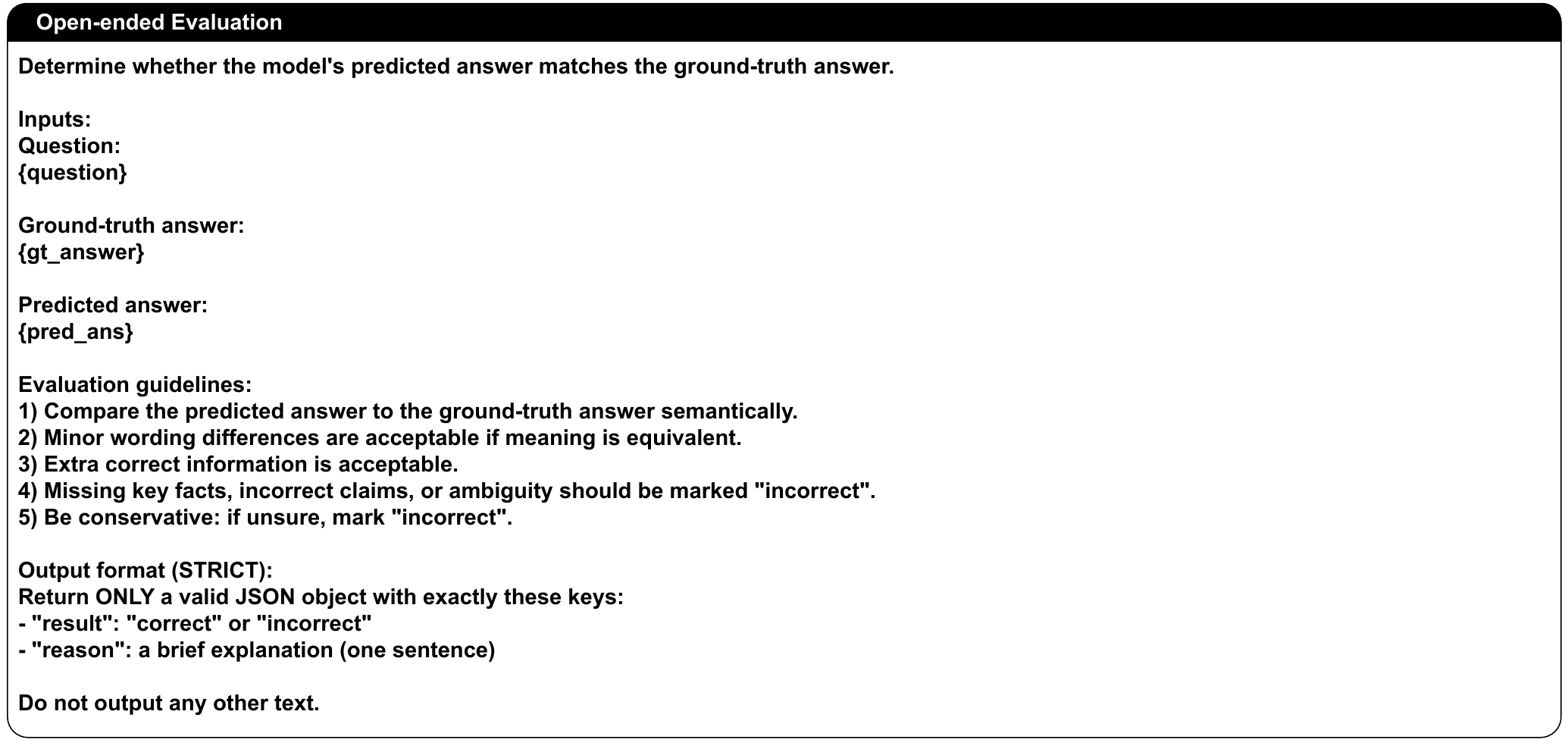}
  \caption{Prompt for Evaluation of Open-ended Type Questions} 
  \label{fig:open_prompt}
\end{figure*}

Evaluation of the model performance can be done in 2 ways depending on the nature of the question.  
\begin{itemize}
    \item Case 1: Multiple Choice Questions: if the questions are multiple choice, evaluation needs to compare the chosen option with the index of the ground truth option, if the model outputs in that specific format. More often than not, the model does not adhere to this format. So either the evaluation needs to compare the predicted option (ideal case), or the option needs to be extracted from the text output by the model. Sometimes, the option may not be present in the text but the answer is. In that case the evaluation is between the context of the predicted output and the ground truth. Thus we pass the ground truth and option number and the predicted answer along with the question and options to GPT-4.1-mini and ask it to score binarily. If the answer (or context) matches the ground truth, it is given the score of 1, otherwise 0. 
    \item Case 2: Open Ended Questions: if the questions do not have predefined options, the output text is very less likely to match the ground truth answer. In that case, understanding the context becomes necessary to evaluate the validity of the prediction. In this case, we pass only the ground truth and the predicted answer along with the question to GPT-4.1-mini. There are no options or ground truth option number in this case. If the output text and the ground truth have the same meaning, it is given the score of 1, otherwise 0. 
\end{itemize}
The prompts for both cases are shown in Figs. \ref{fig:mcq_prompt} and \ref{fig:open_prompt} respectively.

\section{Dataset Details}
\label{sec:data-det}

\subsection{DeepLesion}
\label{sec:deeplesion}
The DeepLesion\cite{yan2018deeplesion} dataset is a large-scale, clinically derived collection of CT scans designed for automated lesion detection and analysis. It contains over 32,000 axial CT slices from more than 4,500 patients, encompassing approximately 32,735 lesions across a variety of organs, including the lung, liver, bone, soft tissue, and lymph nodes. Each lesion is annotated with a 2D bounding box on the slice where it is most visible, along with RECIST diameters, providing standardized size measurements commonly used in radiology. The dataset exhibits significant heterogeneity in imaging protocols, slice thickness (typically 1–5 mm), and lesion size, reflecting the diversity of real-world clinical data. While primarily designed for detection tasks, the bounding box and size annotations also support weakly supervised segmentation and lesion progression analysis. Notably, the dataset does not include explicit malignancy labels, and only lesions visible in the annotated slices are labeled, making it a challenging benchmark for deep learning models in medical imaging. 

In total there are 4831 test images which are slices of CT scans in which the lesion is present. The lesions are from 8 anatomical parts:
\begin{enumerate}
    \item bone
    \item abdomen
    \item mediastinum
    \item liver
    \item lung
    \item kidney
    \item soft tissue
    \item pelvis
\end{enumerate}

To generate data fit for our framework, we formulate each sample into a multiple choice question. The options are the aforementioned coarse labels, while the question is randomly chosen out of these options:
\begin{itemize}
    \item What category best describes the lesion shown in this image?
    \item Which anatomical region does the lesion in this image most likely originate from?
    \item Based on the image, what type of lesion is being demonstrated?
    \item How would you classify the lesion visible in this image?
    \item The lesion shown here is most consistent with involvement of which anatomical structure?
    \item From the imaging appearance, what is the primary site of this lesion?
    \item What type of lesion is depicted in the provided image?
    \item Which body region best characterizes the lesion seen in this image?
    \item Considering the imaging findings, what is the lesion type?
    \item The lesion in this image most likely belongs to which anatomical category?
\end{itemize}

The images from the dataset are losslessly compressed 16-bit png images. We need to subtract 32768 from the 16-bit pixel intensities to obtain the original Hounsfield unit (HU) values. After that the images need to be clipped and normalized to obtain the model and human readable uint8 format. 

\subsection{Examples from Vision Datasets}
\label{sec:vis-ex}
\begin{figure*}[h]
  \centering
  \includegraphics[width =\linewidth]{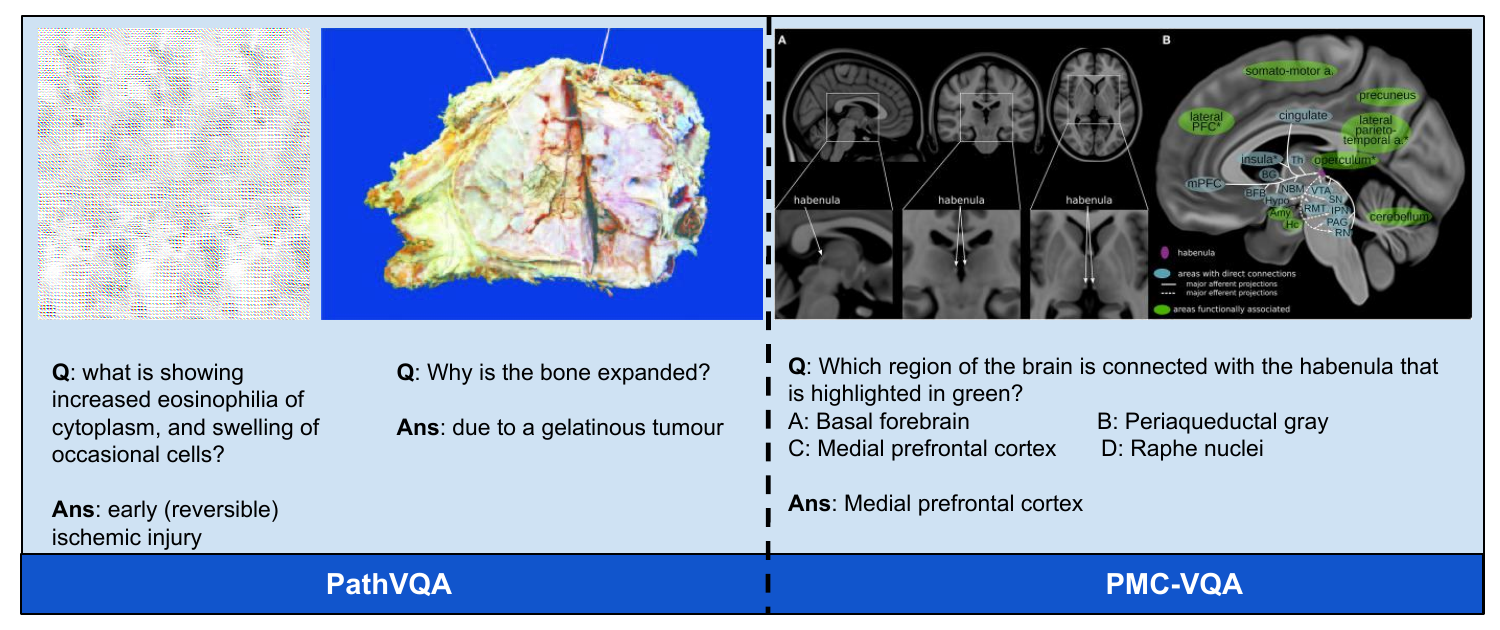}
  \caption{Example QA pairs from Vision Datasets PathVQA and PMC-VQA} 
  \label{fig:ex}
\end{figure*}

The other vision datasets used are PMC-VQA\cite{zhang2024development} amd PathVQA\cite{he2020pathvqa}. Both datasets are general medical VQA datasets, but with some differences. PMC-VQA is a very general dataset consisting of various image types, diseases and disorders, whereas PathVQA is pathology focused. Also PMC-VQA is a multiple choice question with 4 options while PathVQA is open-ended. A few examples from both datasets are shown in Fig. \ref{fig:ex}.

\section{Additional Results on a dataset focusing on a single anatomical feature}
\label{sec:chestxray}
NIH Chestxray8\cite{wang2017chestx} dataset is a large publicly available dataset of frontal-view chest X-ray images designed to facilitate automated thoracic disease detection and classification. It contains over 112,000 X-ray images from more than 30,000 unique patients, annotated with 14 common thoracic pathologies: 
\begin{itemize}
    \item Atelectasis
    \item Consolidation
    \item Infiltration
    \item Pneumothorax
    \item Edema
    \item Emphysema
    \item Fibrosis
    \item Effusion
    \item Pneumonia
    \item Pleural Thickening
    \item Cardiomegaly
    \item Nodule
    \item Mass
    \item Hernia
\end{itemize}
along with a “no findings” category. Each image is associated with image-level labels extracted using natural language processing from radiology reports, and a subset of images includes bounding box annotations for localized pathologies. The dataset exhibits significant variation in patient demographics, imaging equipment, and acquisition settings, reflecting real-world clinical heterogeneity. ChestXray8 has been widely used as a benchmark for multi-label classification, weakly supervised localization, and disease detection in medical imaging.

To convert the dataset into a suitable format for our framework, we follow a similar procedure to DeepLesion (Section \ref{sec:deeplesion}) to make each sample into a multiple choice question with 15 options. This dataset is particularly large with a otal of 17,853 samples left after removing samples with multiple pathologies. We train our router for this dataset as well. As the anatomy features are limited for this dataset, so is the pool of specialists. Despite this restriction, our framework improves significantly upon the GPT-4.1-mini baseline as the baseline gets 10.92\% accuracy while our framework achieves 18.15\%, which demonstrates the robustness of our model to limited anatomy data.


\end{document}